\documentclass[%
 reprint,
superscriptaddress,
 amsmath,amssymb,
 aps,
prb,
]{revtex4-2}

\usepackage{graphicx, color}
\usepackage{dcolumn}
\usepackage{bm}
\usepackage{hyperref}
\hypersetup{
 citecolor=blue,
 colorlinks = true,
 urlcolor = blue
}
\usepackage[normalem]{ulem}
\usepackage{physics}
\allowdisplaybreaks[1]

\begin{document}

\title{Quantum transport of a spin-1 chiral fermion}

\author{Risako Kikuchi}
 \affiliation{Department of Physics, Nagoya University, Nagoya 464-8602, Japan}
\author{Takumi Funato}%
\affiliation{Center for Spintronics Research Network, Keio University, Yokohama 223-8522, Japan}
	\affiliation{%
		Kavli Institute for Theoretical Sciences, University of Chinese Academy of Sciences, Beijing, 100190, China
	}
\author{Ai Yamakage}%
\affiliation{Department of Physics, Nagoya University, Nagoya 464-8602, Japan}

\date{\today}

\begin{abstract}
We theoretically study the quantum transport in a three-dimensional spin-1 chiral fermion system in the presence of impurity scattering. Within the self-consistent Born approximation, we find peak structure of the density of states and significant suppression of electrical conductivity around the zero energy. The zero-energy conductivity depends less on impurity concentration, unlike a Weyl fermion. 
These properties originate from the flat band structure of spin-1 chiral fermion.
\end{abstract}

\maketitle

\section{\label{sec:level1}Introduction}

A variety of chiral quasiparticles beyond Weyl and Dirac fermions can emerge in crystalline material, protected by symmetry and topological charges \cite{Beyond2016}. 
Spin-1 chiral fermion is a prominent realization of such an exotic particle, which hosts the trivial band with nearly flat dispersion, called flat band, in addition to the Dirac cone. Angle-Resolved Photo Emission Spectroscopy (ARPES) measurements have observed a spin-1 chiral fermion in a chiral crystal CoSi \cite{Tang2017-kk,Takane2019, Sanchez2019-by, Rao2019-ts, Yuan2019-hz}. 
A two-dimensional (approximate) version of spin-1 chiral fermion has also been predicted \cite{Shen2010} in the context of the $\alpha$--T$_3$ lattices.
Quantum phenomena, including spin transport \cite{Tang21} and optical responses \cite{Flicker18, Sanchez-Martinez2019-ek, Habe2019-ss, Xu2020-zb, Ni2020-ze, Ni2021-eo, Kaushik2021-hx, Dey2022-th}, can be anomalous, stemming from the peculiar electronic states of a spin-1 fermion. Research has also spilled over into quadratic dispersion with the spin-1 structure \cite{Nandy2019-qw, Chen2021-dv, Pal2022-aw}.
A fundamental example of such phenomena is the electrical conductivity of quantum transport.

Chiral fermions of Dirac and Weyl fermions exhibit peculiar transport properties on zero energy \cite{Hosur2013-ck, Armitage2018-sv, Gorbar2020-mi}. 
For two-dimensional massless Dirac fermions, representing low-energy electrons in graphene, the electrical conductivity remains finite and independent of the relaxation time at their gapless point, though the density of states vanishes \cite{Fradkin1986-sc, Ludwig1994-ln, Shon1998-ke, Ziegler1998-ar, Tworzydlo2006-hz, Katsnelson2006-yd, Noro2010-cm}. 
Weyl fermions in the three spatial dimensions, on the other hand, show a specific quantum transport phenomenon strongly depending on the impurity concentration at the Weyl point located on the zero energy \cite{0minato2014, Kobayashi2014, Nandkishore2014-vz, Ominato2015-um, Ominato2016-jl}. Interestingly, a finite(long)-range disorder potential (does not) gives rise to the semimetal--metal transition.
These studies imply that the quantum transport phenomena on the zero energy possibly depend on the spatial dimension, disorder type, and symmetry/topology of chiral fermions. 
Therefore, a spin-1 chiral fermion, characterized by the flat band, monopole charge 2, and orthogonal symmetry class, is expected to show a distinct quantum transport character. 
Recent years have seen remarkable progress in our understanding of two-dimensional spin-1 systems \cite{Vigh2013, Hausler2015-lp, Yang2019-oj, Burgos2022-rl}. 
Thus, we can address quantum transport phenomena in three-dimensional spin-1 systems.

In this work, we theoretically analyze the quantum transport of three-dimensional spin-1 chiral fermion under impurity potentials, based on the Boltzmann transport theory and the self-consistent Born approximation (SCBA) in linear response theory. 
As a result, we find some characteristic phenomena near the zero energy by SCBA intrinsic to the spin-1 nature. 
There is a peak structure of the density of states, which originates from a flat band of the spin-1 fermion, depending on the impurity concentration. 
We also find significant suppression of conductivity near the zero energy. 
This phenomenon of low-energy transport is caused by the vanishing group velocity of the flat band and the interband effect between the flat band and the Dirac cone.

The paper is organized as follows. 
Our model for a spin-1 chiral fermion in the presence of impurity is introduced in Sec.~\ref{model}.
The conductivity is calculated within the Boltzmann theory in Sec.~\ref{Boltzmann} and the SCBA with the current vertex correction in Sec.~\ref{linear}. 
The self-consistent equation for self energy and the Bethe-Salpeter equation for vertex correction are explicitly derived.
Comparison of our results to related works and some remarks are discussed in Sec.~\ref{discussion}. 
And Sec.~\ref{conclusion} summarizes this work. 

\section{model}
\label{model}
We consider a three-dimensional spin-1 chiral fermion system, described by
\begin{eqnarray}
\hat{\mathcal{H}}=\hbar v\hat{\bm{S}}\cdot \bm{k},
\end{eqnarray}
where $\bm k$ is the electron wavenumber and $v$ is the Fermi velocity. $\bm{S}=(\hat{S}_{x},\hat{S}_{y},\hat{S}_{z})$ are the spin operator: 
\begin{align}
\hat{S}_{x} &=
\pmqty{
0 & i & 0\\
-i & 0 & 0\\
0 & 0 & 0
},
\\
\hat{S}_{y}&=
\pmqty{
0 & 0 & -i\\
0 & 0 & 0\\
i & 0 & 0
},
 \\
\hat{S}_{z} &=
\begin{pmatrix}
0 & 0 & 0\\
0 & 0 & i\\
0 & -i & 0\\
\end{pmatrix}.
\end{align}
The eigenenergy is given by 
\begin{align}
 \epsilon_{\lambda,\bm{k}}=\hbar v\lambda k,
\end{align}
where $\lambda$ is the label for the conduction band ($\lambda = 1$), the flat band ($\lambda = 0$), and the valence band ($\lambda = -1$).

We assume two impurity potentials, the Gaussian and delta function potentials, which represent finite-range and short-range disorders, respectively.
Note that the length scale of impurity may strongly affect the transport property of chiral fermion systems, as seen in graphene.
The Gaussian potential is defined by
\begin{eqnarray}
\label{gauss}
U(\bm{r}) &=&  \frac{\pm u_0}{(\sqrt{\pi}d_0)^3}\exp(-\frac{r^2}{d_0^2}),
\end{eqnarray}
where $d_0$ is the characteristic length scale and $\pm u_0$ is the strength of the impurity potential. 
The sign $\pm$ means to assume that the numbers of positive and negative valued impurities are the same, so the Fermi level is fixed, irrelevant to the impurity concentration.
The delta function potential is defined by
\begin{eqnarray}
\label{delta}
U(\bm{r})=\pm u_0\delta(\bm{r}),
\end{eqnarray}
which corresponds to the short-range limit ($d_0 \to 0$) of the Gaussian potential. Their Fourier transforms are obtained to be
\begin{align}
 u(\bm{k}) =  \pm u_0\exp(-\frac{k^2}{q_0^2}),
 \label{u_Gauss}
\end{align}
with $q_0 = 2/d_0$ and $u(\bm{k})=\pm u_0$, respectively. 
To avoid the ultraviolet divergence, we need to regularize the delta function potential with a hard cutoff as
\begin{align}
u(\bm{k})= 
 \pm u_0 \theta(q_0-k),
 \label{u_delta}
\end{align}
where $q_0$ is the inverse of the length that characterizes the impurity potential.

Isotropic disorder potentials are characterized by the moment of scattering angle as
\begin{align}
V_n^2(k,k') = 2\pi\int_{-1}^1 d(\cos\theta_{\bm{kk'}})|u(\bm{k}-\bm{k'})|^2\cos^n\theta_{\bm{kk'}},
\label{V}
\end{align}
where $\theta_{\bm{kk'}}$ is the angle between $\bm{k}$ and $\bm{k'}$. We also define a parameter characterizing the scattering strength
\begin{eqnarray}
W = \frac{q_0n_{\text{i}}u_0^2}{\hbar^2 v^2},
\end{eqnarray}
where $n_{\text{i}}$ is the number of scatterers per unit volume.

\section{\label{Boltzmann}Boltzmann transport theory}
We calculate the conductivity from the Boltzmann transport theory. 
The scattering probability $W_{\lambda'\bm{k}',\lambda \bm{k}}$ is given by the Fermi's golden rule as
\begin{align}
W_{\lambda'\bm{k'}, \lambda\bm{k}} = \frac{2\pi}{\hbar}n_{\text{i}}|\bra{\lambda',\bm{k'}}U\ket{\lambda,\bm{k}}|^2\delta(\epsilon_{\lambda',\bm{k'}}-\epsilon_{\lambda,\bm{k}}).
\end{align}
The transport relaxation time $\tau_{\mathrm{tr}}$ is defined by
\begin{align}
\frac{1}{\tau_{\rm{tr}}(\epsilon_{\lambda,\bm{k}})} = \sum_{\lambda'}\int\frac{d\bm{k'}}{(2\pi)^3}(1-\cos\theta_{\bm{k'}\bm{k}})W_{\lambda'\bm{k'}, \lambda\bm{k}}.
\end{align}
The density of states of the linear dispersion ($\lambda = \pm 1$) in the clean limit is given by 
\begin{eqnarray}
D_0(\epsilon)=\frac{\epsilon^2}{2\pi^2(\hbar v)^3}
 \qfor \epsilon \neq 0,
\label{dos0}
\end{eqnarray}
and that of the flat band is ($\lambda= 0$) described by the delta function and diverges for $\epsilon = 0$.
The conductivity at the zero temperature is obtained to be
\begin{eqnarray}
\sigma_B(\epsilon)&=& \frac{e^2v^2}{3}D_0(\epsilon)\tau_{\mathrm{tr}}(\epsilon).\label{conductivity_B}
\end{eqnarray}
The transport relaxation time is written as
\begin{align}
\frac{1}{\tau_{\mathrm{tr}}(\epsilon)}
&=\frac{n_{\text{i}}}{2\pi\hbar^2 v}\int_{-1}^1\int_{0}^{\infty} dk'd(\cos\theta_{\bm{kk'}})k'^2(1-\cos\theta_{\bm{kk'}})
\nonumber\\&\quad\times
\frac{(\cos\theta_{\boldsymbol{kk}'}+1)^2}{4}\delta(k-k')|u(\bm{k}-\bm{k'})|^2
\notag\\ & \hspace{-1em}
=\frac{n_{\text{i}}\epsilon^2}{4(2\pi)^2\hbar(\hbar v)^3}
\bigl[V_0^2(\epsilon/\hbar v,\epsilon/\hbar v)+V_1^2(\epsilon/\hbar v,\epsilon/\hbar v)
\notag\\&
-V_2^2(\epsilon/\hbar v,\epsilon/\hbar v)-V_3^2(\epsilon/\hbar v,\epsilon/\hbar v)
\bigr],
\label{conductivity_B_spin1}
\end{align}
where the momentum is located on the Fermi surface, $k=\epsilon/\hbar v$. 

\begin{figure}
\includegraphics[width=7cm]{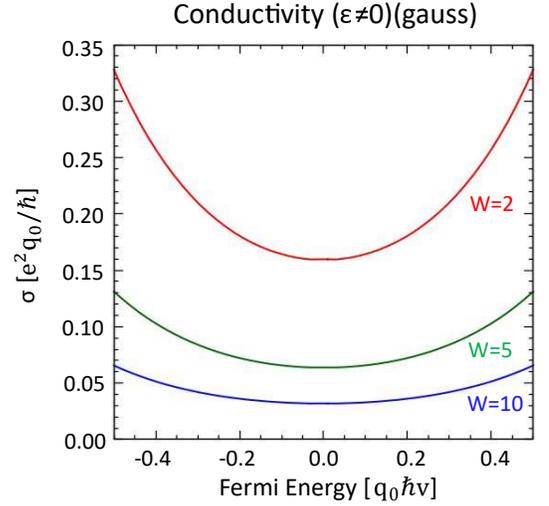}
\caption{(Color online) Electrical conductivity of a spin-1 fermion for $W=2$ (red line), $W=5$ (green line), and $W=10$ (blue line), derived by the Boltzmann equation with the Gaussian potential.}
\label{bol}
\end{figure}
As a result, we find the conductivity 
\begin{eqnarray}
\sigma_B(\epsilon)
&=& \frac{e^2v^2\hbar}{\pi n_{\text{i}}u_0^2},\label{C-Bol-d}
\end{eqnarray}
for the delta function potential with $q_0 \to \infty$. And
\begin{eqnarray}
\sigma_B(\epsilon)
&=& \frac{8}{3\pi}\dfrac{e^2}{\hbar}\dfrac{1}{n_{\text{i}}u_0^2}h\left(\dfrac{\epsilon}{\hbar vq_0}\right),\label{C-Bol-g}
\end{eqnarray}
for the Gaussian potential, where we define
\begin{eqnarray}
h(x)=\dfrac{64x^8}{32x^4-16x^2+3-(8x^2+3)\exp(-8x^2)}.
\end{eqnarray}
Figure~\ref{bol} shows the conductivity for the Gaussian potential, which is a monotonically  increasing function of the Fermi energy and takes a finite value at the zero energy 
\begin{eqnarray}
\sigma_B(\epsilon\rightarrow 0) = \frac{e^2v^2\hbar}{\pi n_{\text{i}}u_0^2},
\end{eqnarray}
which coincides with that of the delta function potential.
The Fermi-energy dependence of conductivity stems from the wavenumber dependence of the potential. The delta and Gaussian-function potentials are constant and exponentially decreasing functions of the wavenumber. Therefore the conductivity is constant and an increasing function for the delta and Gaussian potentials, respectively.

\section{\label{linear}The linear response theory (SCBA)}
Next, we calculate the density of states and conductivity in a self-consistent manner to consider the effect of level broadening induced by impurity.

\subsection{Formulation}
Assuming that the impurity distribution is uniformly random, the impurity-averaged Green function is given by 
\begin{eqnarray}
\hat{G}(\bm{k},\epsilon +is0) 
&=& \frac{1}{\epsilon \hat{S}_0-\hbar vk\hat{\bm{S}}\cdot\bm{n}-\hat{\Sigma}(\bm{k},\epsilon+is0)}, 
\label{green function}
\end{eqnarray}
where $\bm{n}=\bm{k}/k$ is the unit vector and $\hat{S}_0$ is the identity matrix. 
The sign $s$ refers to the retarded ($s=1$) and advanced ($s=-1$) Green's functions.
The self-consistent equation for the self-energy is written as
\begin{align}
\hat{\Sigma}(\bm{k},\epsilon+is0)
 = \int\frac{d\bm{k'}}{(2\pi)^3}n_{\text{i}}|u(\bm{k}-\bm{k'})|^2\hat{G}(\bm{k'},\epsilon+is0).
 \label{self energy}
\end{align}
The density of states per unit volume is calculated as
\begin{align}
D(\epsilon) = -\frac{1}{\pi}\Im\int\frac{d\bm{k}}{(2\pi)^3}\Tr\hat{G}(\bm{k},\epsilon+i0).
\label{dos}
\end{align}
The conductivity by the Kubo formula is written as
\begin{align}
\sigma(\epsilon) &= -\frac{\hbar e^2 v^2}{4\pi}\sum _{s,s'=\pm1}ss'\int\frac{d\bm{k'}}{(2\pi)^3}\text{Tr} [\hat{S}_x\hat{G}(\bm{k'},\epsilon+is0)\nonumber\\
& \quad
\times\hat{J}_x(\bm{k'},\epsilon+is0,\epsilon+is'0)\hat{G}(\bm{k'},\epsilon+is'0)],\label{conductivity}
\end{align}
where $\hat{J}_x(\bm{k},\epsilon,\epsilon')$ is the current density flowing in the $x$ direction reinforced with the vertex correction and is determined to be a solution of the following Bethe-Salpeter equation
\begin{align}
\hat{J}_x(\bm{k},\epsilon,\epsilon') &=
\hat{S}_x + \int\frac{d\bm{k'}}{(2\pi)^3}n_{\text{i}}|u(\bm{k}-\bm{k'})|^2\hat{G}(\bm{k'},\epsilon)\nonumber\\
&\quad\times
\hat{J}_x(\bm{k'},\epsilon,\epsilon')\hat{G}(\bm{k'},\epsilon').\label{Bethe}
\end{align}

Since $(\hat{\bm{S}}\cdot\bm{n})^3=(\hat{\bm{S}}\cdot\bm{n})$ for the spin--1 representation matrices, the self energy is expressed as
\begin{align}
\hat{\Sigma}(\bm{k},\epsilon)
 &= \Sigma_1(k,\epsilon)\hat{S}_0
 +\Sigma_2(k,\epsilon)(\hat{\bm{S}}\cdot\bm{n})+\Sigma_3(k,\epsilon)(\hat{\bm{S}}\cdot\bm{n})^2.\label{self energy2}
\end{align}
Using the above expansion, Eq.~(\ref{green function}) is rewritten as
\begin{align}
\hat{G}(\bm{k},\epsilon)
&= \frac{1}{X(k,\epsilon) \hat{S}_0+Y(k,\epsilon) \hat{\bm{S}}\cdot\bm{n}+Z(k,\epsilon) (\hat{\bm{S}}\cdot\bm{n})^2}
\notag\\
&= x(k,\epsilon) \hat{S}_0+y(k,\epsilon) (\hat{\bm{S}}\cdot\bm{n})+z(k,\epsilon) (\hat{\bm{S}}\cdot\bm{n})^2\label{green function2},
\end{align}
where
\begin{align}
X(k,\epsilon)&=\epsilon-\Sigma_1(k,\epsilon),
\\
Y(k,\epsilon)&=-\hbar vk-\Sigma_2(k,\epsilon),
\\
Z(k,\epsilon)&=-\Sigma_3(k,\epsilon),
\end{align}
and
\begin{align}
x(k,\epsilon)&=\frac{1}{X(k,\epsilon)},
\\
y(k,\epsilon)&=-\frac{Y(k,\epsilon)}{(X(k,\epsilon)+Z(k,\epsilon))^2-Y(k,\epsilon)^2},
\\
z(k,\epsilon)&=\frac{Y(k,\epsilon)^2-Z(k,\epsilon)(X(k,\epsilon)+Z(k,\epsilon))}{((X(k,\epsilon)+Z(k,\epsilon))^2-Y(k,\epsilon)^2)X(k,\epsilon)}.
\end{align}

Here we reduce the above expressions for the self energy to a form more convenient to solve the self-consistent equation.
Substituting Eq.~(\ref{green function2}) into Eq.~(\ref{self energy}), we get 
\begin{align}
	&
\hat{\Sigma}(\bm{k},\epsilon+is0)
\notag\\ &
= \hat{S}_0\int\frac{k'^2dk'}{(2\pi)^3}n_{\text{i}}\left[V_0^2(k,k')x(k',\epsilon+is0)\right.\nonumber\\
& \hspace{7em}
\left.+(V_0^2(k,k')-V_2^2(k,k'))z(k',\epsilon+is0)\right]\nonumber\\
&\quad
+(\hat{\bm{S}}\cdot\bm{n})\int\frac{k'^2dk'}{(2\pi)^3}n_{\text{i}}V_1^2(k,k')y(k',\epsilon+is0)
\nonumber\\
&\quad
+(\hat{\bm{S}}\cdot\bm{n})^2\int\frac{k'^2dk'}{(2\pi)^3}n_{\text{i}}^2\left(\frac{3}{2}V_2^2(k,k')-\frac{1}{2}V_0^2(k,k')\right)\nonumber\\
&\hspace{10em}\times 
z(k',\epsilon+is0),
\end{align}
with the help of useful relations shown in Appendix \ref{integral}.
Comparing this with Eq.~(\ref{self energy2}), the self-consistent equation is decomposed into the three equations as
\begin{align}
	&
\Sigma_1(k,\epsilon+is0)
 =\int\frac{k'^2dk'}{(2\pi)^3}n_{\text{i}}
 \bigl[V_0^2(k,k')x(k',\epsilon+is0)
\nonumber\\& \hspace{4em} +
(V_0^2(k,k')-V_2^2(k,k'))z(k',\epsilon+is0) \bigr],
\label{self1}
\\ &
\Sigma_2(k,\epsilon+is0)
=\int\frac{k'^2dk'}{(2\pi)^3}n_{\text{i}}V_1^2(k,k')y(k',\epsilon+is0),
\label{self2}
\\&
\Sigma_3(k,\epsilon+is0) = \int\frac{k'^2dk'}{(2\pi)^3}n_{\text{i}}\left(\frac{3}{2}V_2^2(k,k')-\frac{1}{2}V_0^2(k,k')\right)\nonumber\\&
\hspace{12em}
\times z(k',\epsilon+is0).
\label{self3}
\end{align}
Substituting the self energy into Eq.~(\ref{dos}), the density of states is written as
\begin{align}
&D(\epsilon)
= -\frac{1}{\pi}\Im\int\frac{d\bm{k}}{(2\pi)^3}\left(\frac{1}{X(k,\epsilon+i0)} \right.
\nonumber\\&\quad
+\frac{1}{X(k,\epsilon+i0)+Y(k,\epsilon+i0)+Z(k,\epsilon+i0)}\nonumber\\
&\quad\left.
+\frac{1}{X(k,\epsilon+i0)-Y(k,\epsilon+i0)+Z(k,\epsilon+i0)}\right).
\end{align}

In addition, the Bethe-Salpeter equation is simplified into an easier form. 
The current vertex $\hat{J}_x(\bm{k},\epsilon,\epsilon')$ is expanded to eight terms as
\begin{align}
\hat{J}_x(\bm{k},\epsilon,\epsilon')
&=\hat{S}_xJ_0(k,\epsilon,\epsilon')+n_x(\hat{\bm{S}}\cdot\bm{n})^2J_1(k,\epsilon,\epsilon')\nonumber\\
&+n_x(\hat{\bm{S}}\cdot\bm{n})J_2(k,\epsilon,\epsilon')+(\hat{\bm{S}}\cdot\bm{n})^2\hat{S}_xJ_3(k,\epsilon,\epsilon')\nonumber\\
&+\hat{S}_x(\hat{\bm{S}}\cdot\bm{n})^2J_4(k,\epsilon,\epsilon')+(\hat{\bm{S}}\cdot\bm{n})\hat{S}_xJ_5(k,\epsilon,\epsilon')\nonumber\\
&+\hat{S}_x(\hat{\bm{S}}\cdot\bm{n})J_6(k,\epsilon,\epsilon')+n_x\hat{S}_0J_7(k,\epsilon,\epsilon')\label{J},
\end{align}
by using Eqs.~(\ref{int1})--(\ref{int2}) for Eq.~(\ref{Bethe})~(see Appendix~\ref{integral}).
The matrix-form Bethe-Salpeter equation reduces to the eight equations (see Appendix~\ref{j}) for the expansion coefficients $J_0$--$J_7$. 
Substituting them into Eq.~(\ref{conductivity}), the conductivity is written as
\begin{widetext}
\begin{eqnarray}
&&\sigma(\epsilon)= \frac{2\hbar e^2 v^2}{3}\int_0^\infty \frac{k'^2dk'}{(2\pi)^3}\mbox{Re}\left[-\frac{J_0^{++}+J_1^{++}+J_2^{++}+J_3^{++}+J_4^{++}+J_5^{++}+J_6^{++}+J_7^{++}}{(X+Y+Z)^2}\right.\nonumber\\
&&-\frac{J_0^{++}-J_1^{++}+J_2^{++}+J_3^{++}+J_4^{++}-J_5^{++}-J_6^{++}-J_7^{++}}{(X-Y+Z)^2}-\frac{2J_0^{++}+J_3^{++}+J_4^{++}+J_5^{++}+J_6^{++}}{X(X+Y+Z)}\nonumber\\
&&-\frac{2J_0^{++}+J_3^{++}+J_4^{++}-J_5^{++}-J_6^{++}}{X(X-Y+Z)}+\frac{J_0^{+-}+J_1^{+-}+J_2^{+-}+J_3^{+-}+J_4^{+-}+J_5^{+-}+J_6^{+-}+J_7^{+-}}{|X+Y+Z|^2}\nonumber\\
&&+\frac{J_0^{+-}-J_1^{+-}+J_2^{+-}+J_3^{+-}+J_4^{+-}-J_5^{+-}-J_6^{+-}-J_7^{+-}}{|X-Y+Z|^2}+\frac{J_0^{+-}+J_4^{+-}+J_6^{+-}}{X(X^*+Y^*+Z^*)}\nonumber\\
&&+\frac{J_0^{+-}+J_4^{+-}-J_6^{+-}}{X(X^*-Y^*+Z^*)}\left.+\frac{J_0^{+-}+J_3^{+-}+J_5^{+-}}{X^*(X+Y+Z)}+\frac{J_0^{+-}+J_3^{+-}-J_5^{+-}}{X^*(X-Y+Z)}\right],
\end{eqnarray}
\end{widetext}
where $J_i^{ss'}=J_i(k',\epsilon+is0,\epsilon+is'0)$, $X=X(k',\epsilon+i0)$, and so on.

\subsection{\label{calculation}Numerical Calculations}
The self-consistent equations derived above cannot be generally solved in an analytic way, but we can obtain the solution by numerical iteration \cite{Noro2010-cm}.
We discretize the wavenumber as
\begin{align}
dk_j &=  k_c\frac{j}{\sum_{j=1}^{j_{\text{max}}}j}\label{numerical},
\quad
k_j = \frac{1}{2}dk_j + \sum_{j'=1}^{j-1}dk_{j'},
\end{align}
where $j=1,2,...,j_{\text{max}}$ and $k_c$ is the cutoff wavenumber. 
Hereafter, we fix $j_{\text{max}}=500$.

\subsection{\label{result_scba}Density of states}
\begin{figure*}
\includegraphics[width=14cm]{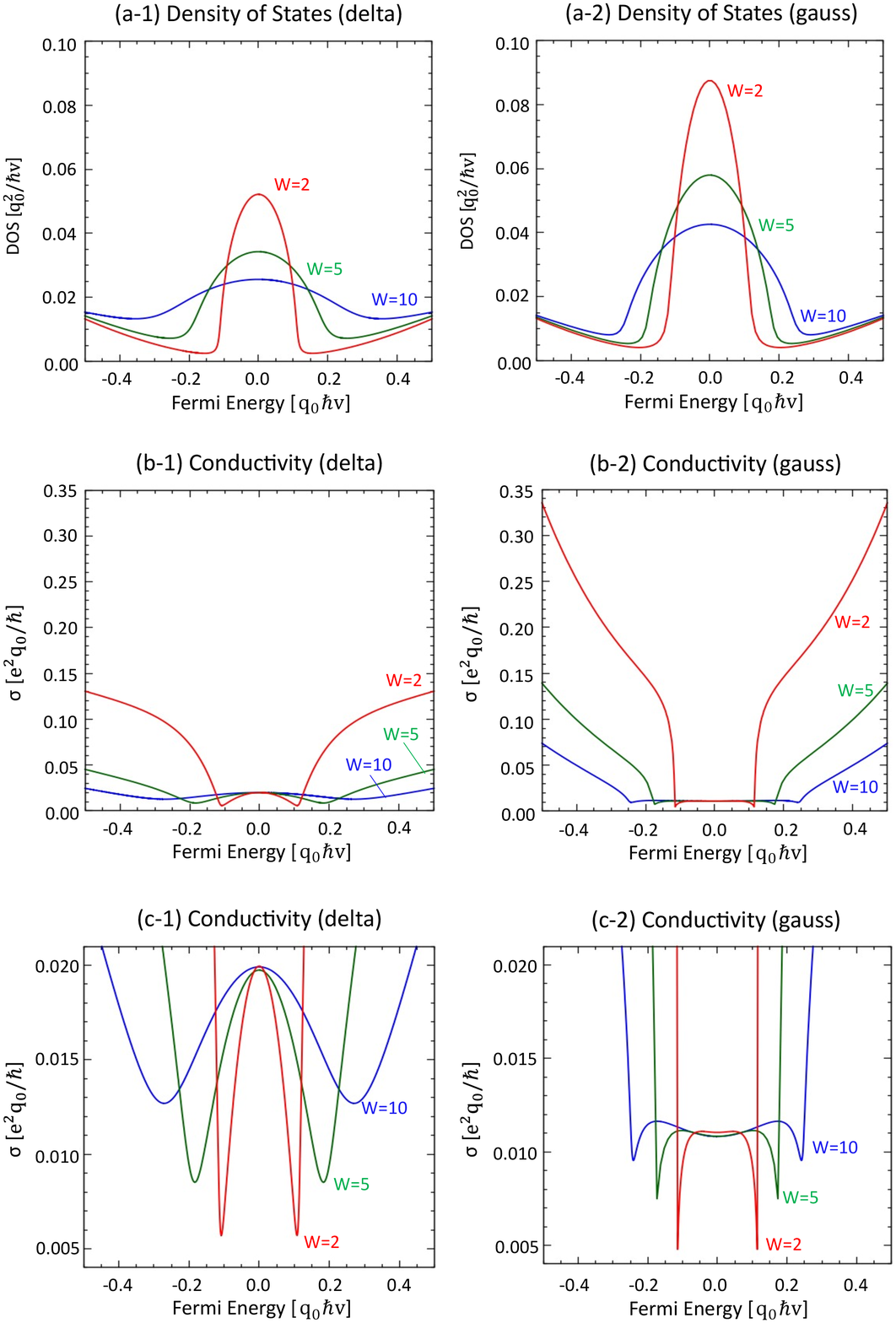}
\caption{(Color online) Quantum transport for $W=2$ (red line), $W=5$ (green line), and $W=10$ (blue line), derived by the SCBA. 
	Density of states for (a-1) the delta function potential and (a-2) Gaussian potential ($k_c=q_0$). 
	Conductivity for (b-1) the delta function potential and (b-2) the Gaussian potential ($k_c=q_0$). 
	(c-1) and (c-2) are the enlargements near the zero energy of (b-1) and (b-2), respectively.}
\label{impurity}
\end{figure*}

The density of states and conductivity are obtained by SCBA. 
We fix $k_c=q_0$ in the following.
Note that the results do not explicitly depend on $q_0$ because the density of states and the conductivity are functions of $\epsilon/(q_0 \hbar v)$ and $W$, and is normalized by $q_0^2/\hbar v$ and $e^2 q_0/\hbar$, respectively.

Figures~\ref{impurity}(a-1) and \ref{impurity}(a-2) show the density of states as a function of the Fermi energy for the delta function and Gaussian potentials, respectively.
We can see a pronounced peak structure around the zero energy.
This peak stems from the flat band at $\epsilon=0$ and is broaden by the impurity potential.
Away from the zero energy, the density of states is approximately proportional to $\epsilon^2$, which comes from the linear dispersion bands and is essentially the same as those in the clean limit.

\subsection{Conductivity}

Figures~\ref{impurity}(b-1) and \ref{impurity}(b-2) show the conductivity as a function of the Fermi energy for the delta function and Gaussian potentials, respectively.
In the high-energy region, for both the delta function/Gaussian potential, the conductivity increases monotonically and reaches nearly the values given in the Boltzmann theory. 
However, the conductivity is significantly suppressed in the vicinity of the zero energy. 
The energy range of the suppressed area is as wide as that of the peak of the density of states. 
This implies that the flat band plays a crucial role in suppressing conductivity. 

In addition, the suppressed conductivity exhibits a behavior specific to the length scale of potential. 
Figures~\ref{impurity}(c-1)~and~\ref{impurity}(c-2) show enlarged views of the suppressed area. 
The conductivity for the delta function potential [Fig.~2(c-1)] shows a small peak, whereas that for the Gaussian potential [Fig.~2(c-2)] is nearly constant with a tiny dip on the zero energy.
This difference is due to the vertex corrections $J_1-J_7$ in Eq.~(\ref{J}).
We clarified that the suppressed conductivity for the Gaussian potential without vertex correction ($J_0=1$, $J_1=J_2=...=J_7=0$) makes a small peak, as for the delta function potential. Namely, the vertex correction plays a vital role in the low-energy regime. 
Unlike the Gaussian potential, the conductivity for the delta function potential produces a small peak and is not flattened by the vertex correction, which is given by $J_0 \ne 0$ and $J_1=J_2=...=J_7=0$.
Note that the suppressed conductivity shows less dependence on $W$ than the unsuppressed area in the higher energy.

In this study, we assume zero temperature. The conductivity at a finite temperature is given by $\int d\epsilon (-\partial f / \partial \epsilon) \sigma(\epsilon)$. 
	The Fermi-energy dependence of the conductivity is smeared by the Fermi distribution $f$. In spin-1 systems, since the conductivity has a large Fermi-energy dependence, the temperature dependence may be nonmonotonic.

\subsection{Interband effect}

	Here we clarify the details of the characteristic behavior near zero energy and its origin. To do so, we decompose the conductivity into intraband and interband contributions, which are defined by
	\begin{align}
		\sigma_{\text{intra}}(\epsilon) &= -\frac{\hbar e^2 v^2}{4\pi}\sum _{s,s'=\pm1}ss'\int\frac{d\bm{k'}}{(2\pi)^3}
		\bigl(S_{cc}G^s_c J^{ss'}_{cc}G^{s'}_c\nonumber\\&+S_{vv} G^s_v J^{ss'}_{vv} G^{s'}_v  \bigr),
		\label{intra}
	\end{align}
	and
	\begin{align}
		\sigma_{\text{inter}}(\epsilon) &= -\frac{\hbar e^2 v^2}{4\pi}\sum _{s,s'=\pm1}ss'\int\frac{d\bm{k'}}{(2\pi)^3}
		\bigl(S_{0c} G^s_c J^{ss'}_{c0} G^{s'}_0\nonumber\\&+S_{c0} G^s_0 J^{ss'}_{0c} G^{s'}_c +S_{0v} G^s_v J^{ss'}_{v0} G^{s'}_0 \nonumber\\&+S_{v0} G^s_0 J^{ss'}_{0v} G^{s'}_v \bigr),
		\label{inter}
	\end{align}
	where the subscripts $c$, $0$, and $v$ denote the conduction, flat, and valence bands in the band basis, respectively.
	They are obtained by diagonalizing the Green's function matrix as
	\begin{eqnarray}
		\hat{U}^{\dagger}\hat{G}(\bm{k},\epsilon+is0)\hat{U}=
		\begin{pmatrix}
			G^s_c & 0 & 0\\
			0 & G^s_0 & 0\\
			0 & 0 & G^s_v\\
		\end{pmatrix}.
	\end{eqnarray}
	In this basis, the velocities $S_x$ and $J_x$ are written as
	\begin{eqnarray}
		\hat{U}^{\dagger}\hat{S}_{x}\hat{U}=
		\begin{pmatrix}
			S_{cc} & S_{c0} & 0\\
			S_{0c} & 0 & S_{0v}\\
			0 & S_{v0} & S_{vv}\\
		\end{pmatrix},
	\end{eqnarray}
	and
	\begin{eqnarray}
		\hat{U}^{\dagger}\hat{J_{x}}(k, \epsilon+is0, \epsilon+is'0)\hat{U}=
		\begin{pmatrix}
			J^{ss'}_{cc} & J^{ss'}_{c0} & 0\\
			J^{ss'}_{0c} & J^{ss'}_{00} & J^{ss'}_{0v}\\
			0 & J^{ss'}_{v0} & J^{ss'}_{vv}\\
		\end{pmatrix}.
	\end{eqnarray}
	The flat band has zero velocity leading to the absence of intraband terms.
	The interband terms between the conduction and valence bands are also zero. 

\begin{figure*}
	\includegraphics[width=17.5cm]{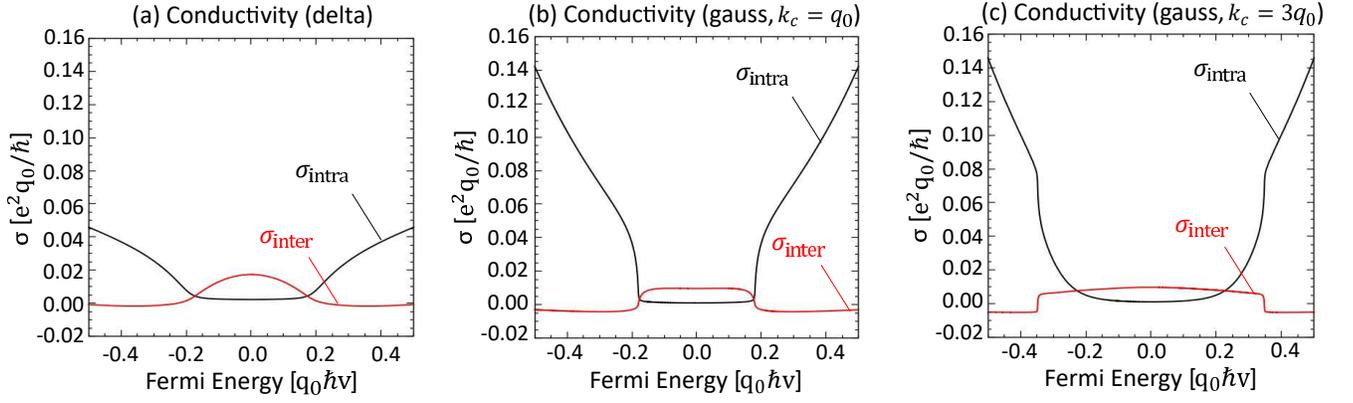}
	\caption{(Color online) The conductivity from the intraband transition of the Dirac cone (black line) and from the interband transition between the Dirac cone and the flat band (red line) for $W=5$ derived by the SCBA. (a) delta function potential, (b) Gaussian potential, $k_c=q_0$ and (c) Gaussian potential, $k_c=3q_0$}
	\label{band}
\end{figure*}
	
		Figure~\ref{band} shows the decomposed conductivities in the band basis. 
		One can see a small peak of the interband term and suppression of the intraband terms near $\epsilon=0$. 
		Namely, the interband conductivity is more dominant than the intraband one.
		This suggests that disorder effects are decisive for the low-energy transport behavior.

\subsection{\label{sec:level2}Cutoff dependence}\label{Cutoff dependence}
\begin{figure*}
\includegraphics[width=17.5cm]{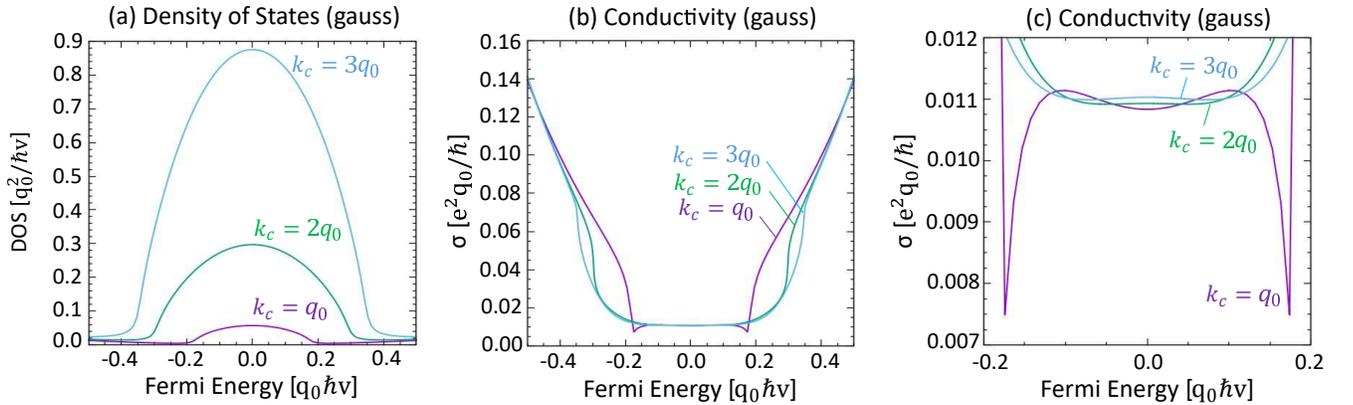}
\caption{(Color online) Quantum transport for $k_c=q_0$ (purple line), $k_c=2q_0$ (green line) and $k_c=3q_0$ (blue line) derived by the SCBA. (Gaussian potential, $W=5$) (a) Density of states, (b) Conductivity and (c) Conductivity near the suppressed area}
\label{kc}
\end{figure*}

Figure~\ref{kc} is the result of varying the value of $k_c$ in Eq.~(\ref{numerical}) for the Gaussian potential.
Note that we fix $W=5$ here. 
One can see that the peak of the density of states shown in Fig.~(\ref{kc}) rapidly increases as $k_c$ increases. 
Our model has an entirely flat band, which is a simple approximation for that in the actual material. 
Namely, the cutoff momentum $k_c$ corresponds to the range over which the band can be approximated as flat. Consequently, as $k_c$ increases, the contribution from the flat band also increases, resulting in a large peak structure in the density of states. 

The conductivity shown in Fig.~\ref{kc}(b) indicates that the energy region showing the suppressed conductivity ($\sigma \sim 0.011 e^2 q_0/\hbar$) becomes wider as $k_c$ increases for the same reason as for the enhancement of the density of states discussed above.
We also find that the conductivity at $\epsilon \sim 0$ does not change significantly with $k_c$ [Fig.~\ref{kc}(c)]. 
As the density of states increases, the number of scattering channels also increases, and as a result, the conductivity is expected to remain nearly constant.

\section{\label{discussion}Discussion}

\subsection{Comparison of results from Boltzmann equation and SCBA}
\begin{figure}
\includegraphics[width=7cm]{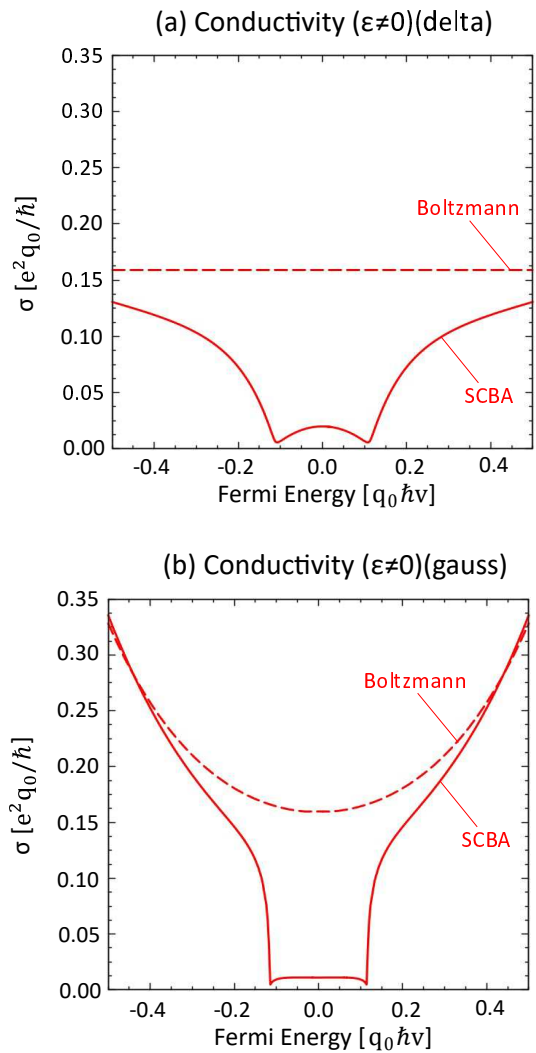}
\caption{(Color online) The conductivity ($W=2$) derived by the Boltzmann equation (dashed line) and the SCBA (solid line) for (a) the delta function potential and (b) Gaussian potential ($k_c=q_0$).}
\label{compare}
\end{figure}

Figure~\ref{compare} compares the conductivity derived from the Boltzmann equation with that derived from SCBA.
The specific behavior of the low-energy region (suppressed conductivity) derived from SCBA cannot be derived from the Boltzmann equation for the following reasons.
The suppressed conductivity originates from the interband effects of the Dirac cone and the flat band broadened by impurity scattering. 
The spectral broadening is not considered in the Boltzmann equation hence the interband effect does not contribute to the conductivity, owing to the energy conservation. 
SCBA is the simplest method to take spectral broadening into account, and it provides a simple way to understand interband conduction phenomena in spin-1 systems.

\subsection{Related studies}


\begin{figure}
\includegraphics[width=7cm]{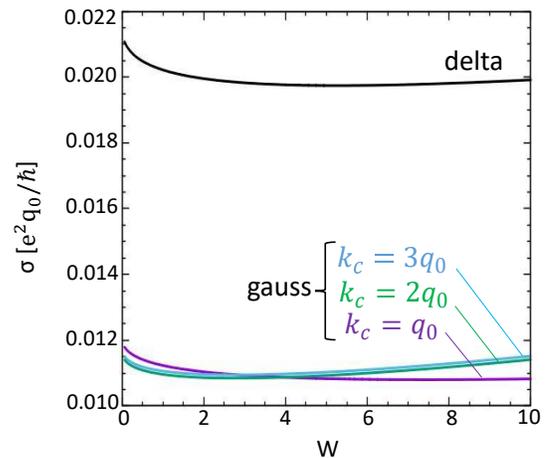}
\caption{(Color online) Conductivity on the zero energy for the delta function potential (black line) and Gaussian potential ($k_c=q_0$:purple line, $k_c=2q_0$:green line and $k_c=3q_0$:blue line) derived by the SCBA.}
\label{epsilon0}
\end{figure}

In a three-dimensional Weyl fermion under the finite range (Gaussian) impurity potential, the conductivity at the Weyl point changes its behavior significantly depending on the scattering strength, indicating the semimetal--metal transition \cite{0minato2014}
.
In the spin-1 fermion system, on the contrary, the conductivity at $\epsilon=0$ has a much smaller dependence on the scattering strength, showing no transition
. 

A spin-1 fermion in two spatial dimension is not protected by symmetry but approximately emerges. 
In the two-dimensional spin-1 fermion with the delta function disorder potential, the density of states has a peak, and the conductivity is suppressed near the zero energy \cite{Vigh2013} like in the three-dimensional system. 
The behavior of the conductivity suppression differs between the 2D and 3D systems, with the 2D system showing a gradual increase as energy approaches zero, whereas the 3D system shows a rapid increase and makes a small peak.
Furthermore, we elucidate a dependence of the length scale of impurity of the conductivity for the three-dimensional system.  
The short-range (delta function) potential makes a peak, while the finite-range (Gaussian) one does not. 
This low-energy spectral property of conductivity can be essential to comparing and understanding experimental results in the future.

\subsection{On experimental realization}

The present study has shown the Fermi-energy dependences of the density of states and the conductivity. Here we comment on relation to experimental realization. 
A direct way to tune the Fermi energy is to use the gating of a thin-film spin-1 fermion material. 
Alternatively, the Fermi energy could be varied by doping the bulk material, although this is not easy to control continuously.
 A chiral crystal CoSi hosts a spin-1 chiral fermion on the Brillouin zone center in the absence of spin-orbit interaction, which would be a good platform to observe the quantum transport of the spin-1 fermion. 
 However, in addition to the spin-1 fermion at the $\Gamma$ point, there is a double Weyl fermion at the $R$ point, and one must take the additional contribution from them into the transport properties. Further research on this is desirable.

\section{\label{conclusion}Conclusion}
This study clarified the quantum transport theory for a spin-1 chiral fermion with disorder potential within the SCBA in the presence of the current vertex correction. 
As a result, we found some characteristic phenomena originated from the flat-band structure near the zero energy. 
One is a peak structure of the density of states.
The other is the suppression of conductivity. 
It has been discussed that 
these apparently contradicting behaviors are reconciled by considering the vanishing group velocity of the flat band. 
In addition, we have found that the interband effect, which is inherent to the spin-1 band structure, significantly contributes to the electrical conductivity.

These results provide the basis for clarifying the quantum transport phenomena of spin-1 fermions. Furthermore, it is suggested that nontrivial impurity effects are latent in chiral fermions beyond Dirac and Weyl fermions, and it will be an interesting problem to clarify quantum transport phenomena in more diverse chiral-fermion systems.

\textit{Note added.} 
We became aware of a very recent work which studies the density of states \cite{hsu2022disorder}, Fermi arcs and surface Berry curvature for the spin-1 chiral fermion in the presence of disorder. We have found the same qualitative behavior of the density of states.

\begin{acknowledgments}
	This work is supported by JSPS KAKENHI for Grants (Grants No. JP20K03835 and No. JP21K20356) and the Sumitomo Foundation (Grant No. 190228).
\end{acknowledgments}

\appendix

\section{Useful relations}\label{integral}

Let $\bm{n}_{\perp 1}$, $\bm{n}_{\perp 2}$, and $\bm{n}$ be the three-dimensional unit vectors that are perpendicular mutually. 
Let $S_x$, $S_y$, and $S_z$ be the $3 \times 3$ spin-1 representation matrices introduced in the main text. 
One finds the following useful relations.
\begin{align}
	&
(\hat{\bm{S}}\cdot\bm{n})^3=(\hat{\bm{S}}\cdot\bm{n}),
\label{UR1}
\\&
(\hat{\bm{S}}\cdot\bm{n})^2\hat{S}_{i}(\hat{\bm{S}}\cdot\bm{n})^2=(\hat{\bm{S}}\cdot\bm{n})\hat{S}_{i}(\hat{\bm{S}}\cdot\bm{n})=n_{i}(\hat{\bm{S}}\cdot\bm{n}),
\\&
(\hat{\bm{S}}\cdot\bm{n}_{\perp 1})^2+(\hat{\bm{S}}\cdot\bm{n}_{\perp 2})^2+(\hat{\bm{S}}\cdot\bm{n})^2=2\hat{S}_0,
\\&
(\hat{\bm{S}}\cdot\bm{n}_{\perp 1})\hat{S}_{i}(\hat{\bm{S}}\cdot\bm{n}_{\perp 1})+(\hat{\bm{S}}\cdot\bm{n}_{\perp 2})\hat{S}_{i}(\hat{\bm{S}}\cdot\bm{n}_{\perp 2})
\nonumber\\&
+(\hat{\bm{S}}\cdot\bm{n})\hat{S}_{i}(\hat{\bm{S}}\cdot\bm{n})=\hat{S}_{i},
\\&
(\hat{\bm{S}}\cdot\bm{n}_{\perp 1})\hat{S}_{i}\hat{S}_{j}(\hat{\bm{S}}\cdot\bm{n}_{\perp 1})+(\hat{\bm{S}}\cdot\bm{n}_{\perp 2})\hat{S}_{i}\hat{S}_{j}(\hat{\bm{S}}\cdot\bm{n}_{\perp 2})
\nonumber\\&
+(\hat{\bm{S}}\cdot\bm{n})\hat{S}_{i}\hat{S}_{j}(\hat{\bm{S}}\cdot\bm{n})=-\hat{S}_{j}\hat{S}_{i}+2\delta_{ij}\hat{S}_0.
\label{UR5}
\end{align}

An arbitrary unit vector $\boldsymbol{n}'$ is expressed as
\begin{align}
	\bm{n'}
	=\bm{n}_{\perp 1}\sin\theta\cos\phi+\bm{n}_{\perp 2}\sin\theta\sin\phi+\bm{n}\cos\theta,
\end{align}
where $\theta$ denotes the angle between $\bm{n}$ and $\bm{n'}$ and $\phi$ denotes the azimuth angle in the $\bm{n}_{\perp1}$-$\bm{n}_{\perp 2}$ plane.

Using Eqs.~(\ref{UR1})--(\ref{UR5}), we can calculate the following integrals as
\begin{align}
	&
\int_0^{2\pi}\int_0^\pi d\theta d\phi|u(\bm{k}-\bm{k'})|^2(\hat{\bm{S}}\cdot\bm{n'})
=(\hat{\bm{S}}\cdot\bm{n})V_1^2(k,k'),\label{int1}
\\&
\int_0^{2\pi}\int_0^\pi d\theta d\phi|u(\bm{k}-\bm{k'})|^2(\hat{\bm{S}}\cdot\bm{n'})^2
\nonumber\\&
=\left(\frac{3}{2}V_2^2(k,k')-\frac{1}{2}V_0^2(k,k')\right)(\hat{\bm{S}}\cdot\bm{n})^2
\nonumber\\&\quad
+\left(V_0^2(k,k')-V_2^2(k,k')\right)\hat{S}_0,
\\&
\int_0^{2\pi}\int_0^\pi d\theta d\phi|u(\bm{k}-\bm{k'})|^2n'_x\hat{S}_0
=V_1^2(k,k')n_x\hat{S}_0,
\\&
\int_0^{2\pi}\int_0^\pi d\theta d\phi|u(\bm{k}-\bm{k'})|^2n'_x(\hat{\bm{S}}\cdot\bm{n'})
\nonumber\\&
=\left(\frac{3}{2}V_2^2(k,k')-\frac{1}{2}V_0^2(k,k')\right)n_x(\hat{\bm{S}}\cdot\bm{n})
\nonumber\\&\quad
+\frac{1}{2}(V_0^2(k,k')-V_2^2(k,k'))\hat{S}_{x},
\\&
\int_0^{2\pi}\int_0^\pi d\theta d\phi|u(\bm{k}-\bm{k'})|^2n'_x(\hat{\bm{S}}\cdot\bm{n'})^2
\nonumber\\&
=\frac{1}{2}(V_1^2(k,k')-V_3^2(k,k'))\hat{S}_{x}(\hat{\bm{S}}\cdot\bm{n})
\nonumber\\&\quad
+\frac{1}{2}(V_1^2(k,k')-V_3^2(k,k'))(\hat{\bm{S}}\cdot\bm{n})\hat{S}_{x}
\nonumber\\&\quad
+(V_1^2(k,k')-V_3^2(k,k'))n_x\hat{S}_0
\nonumber\\&\quad
+\left(\frac{5}{2}V_3^2(k,k')-\frac{3}{2}V_1^2(k,k')\right)n_x(\hat{\bm{S}}\cdot\bm{n})^2.\label{int2}
\end{align}

\section{The Bethe-Salpeter equation}\label{j}
 From Appendix~\ref{integral}, the Bethe-Salpeter equation becomes Eq.~(\ref{vertex}) where $J_i=J_i(k,\epsilon+is0,\epsilon+is'0),J'_i=J_i(k',\epsilon+is0,\epsilon+is'0), V_i^2=V_i^2(k,k'),x=x(k',\epsilon+is0),x'=x(k',\epsilon+is'0)$ and so on.
\begin{widetext}
\begin{eqnarray}
\left( \begin{array}{c}  J_0\\
 J_1\\
 J_2\\
 J_3\\
 J_4\\
 J_5\\
 J_6\\
 J_7 \end{array}\right)
&=&
\left( \begin{array}{c}  1\\
0\\
0\\
0\\
0\\
0\\
0\\ 
0 \end{array}\right)
+
\int \displaystyle\frac{k'^2dk'}{(2\pi)^3}n_{\text{i}}\begin{pmatrix}
V_0^2 & 0 & \frac{1}{2}V_0^2-\frac{1}{2}V_2^2 & V_0^2-V_2^2 & V_0^2-V_2^2 & 0 & 0 & 0 \\
0 & \frac{5}{2}V_3^2-\frac{3}{2}V_1^2 & 0 & 0 & 0 & 0 & 0 & 0 \\
0 & 0 & \frac{3}{2}V_2^2-\frac{1}{2}V_0^2 & 0 & 0 & 0 & 0 & 0 \\
0 & 0 & 0 & \frac{3}{2}V_2^2-\frac{1}{2}V_0^2 & 0 & 0 & 0 & 0 \\
0 & 0 & 0 & 0 & \frac{3}{2}V_2^2-\frac{1}{2}V_0^2 & 0 & 0 & 0 \\
0 & \frac{1}{2}V_1^2-\frac{1}{2}V_3^2 & 0 & 0 & 0 & V_1^2 & 0 & 0 \\
0 & \frac{1}{2}V_1^2-\frac{1}{2}V_3^2 & 0 & 0 & 0 & 0 & V_1^2 & 0 \\
0 & V_1^2-V_3^2 & 0 & 0 & 0 & 0 & 0 & V_1^2 \\
\end{pmatrix}\hat{T}
\left( \begin{array}{c}  J_0'\\
 J_1'\\
 J_2'\\
 J_3'\\
 J_4'\\
 J_5'\\
 J_6'\\
 J_7'\end{array}\right)\label{vertex},\nonumber\\
\end{eqnarray}
\end{widetext}
where the matrix $\hat T$ is defined as
\begin{align}
\hat{T}=\begin{pmatrix}
xx' & 0 & 0 & 0 & 0 & 0 & 0 & 0 \\
T_{01} & T_{11} & T_{21} & T_{31} & T_{41} & T_{51} & T_{61} & T_{71} \\
T_{02} & T_{12} & T_{22} & T_{32} & T_{42} & T_{52} & T_{62} & T_{72} \\
zx' & 0 & 0 & T_{33} & 0 & yx' & 0 & 0 \\
xz' & 0 & 0 & 0 & T_{44} & 0 & xy' & 0 \\
yx' & 0 & 0 & yx' & 0 & T_{55} & 0 & 0 \\
xy' & 0 & 0 & 0 & xy' & 0 & T_{66} & 0 \\
0 & 0  & 0  & 0  & 0  & 0  & 0  & xx' \\
\end{pmatrix}.
\end{align}
Here, the matrix elements $T_{ij}$ in the second line of $\hat T$ are given by
\begin{align}
T_{01}&=yz'+zy',
\\
T_{11}&=xx'+xz'+yy'+zx'+zz',
\\
T_{21}&=xy'+yx'+yz'+zy',
\\
T_{31}&=xy'+yz'+zy',
\\
T_{41}&=yx'+yz'+zy',
\\
T_{51}&=xz'+yy'+zz',
\\
T_{61}&=yy'+zx'+zz',
\\
T_{71}&=xz'+yy'+zx'+zz',
\end{align}
$T_{ij}$ in the third line are given by
\begin{align}
T_{02}&=yy'+zz',
\\
T_{12}&=xy'+yx'+yz'+zy',
\\
T_{22}&=xx'+xz'+yy'+zx'+zz',
\\
T_{32}&=xz'+yy'+zz',
\\
T_{42}&=yy'+zx'+zz',
\\
T_{52}&=xy'+yz'+zy',
\\
T_{62}&=yx'+yz'+zy',
\\
T_{72}&=xy'+yx'+yz'+zy',
\end{align}
and, the others are given by
\begin{align}
T_{33}&=xx'+zx',
\\
T_{44}&=xx'+xz',
\\
T_{55}&= xx'+zx',
\\
T_{66}&=xx'+xz'.
\end{align}

By solving these eight self-consistent equations, $J_0$--$J_7$ are determined.

\bibliography{ref}
\clearpage
\end{document}